\documentclass[a4paper,11pt]{article}
\usepackage{pos}
\usepackage{subcaption}

\title{Thermal QCD phase transition and its scaling window from Wilson twisted mass fermions}
\ShortTitle{Thermal QCD phase transition and its scaling window...}

\author*[a]{A.Yu.~Kotov}
\author[b]{M.P.~Lombardo}
\author[c]{A.~Trunin}

\affiliation[a]{Bogoliubov Laboratory of Theoretical Physics, Joint Institute for Nuclear Research, Dubna, 141980 Russia}
\affiliation[b]{INFN, Sezione di Firenze, 50019 Sesto Fiorentino (FI), Italy}
\affiliation[c]{Samara National Research University, Samara, 443086 Russia}

\emailAdd{kotov.andrey.yu@gmail.com}
\emailAdd{lombardo@fi.infn.it}
\emailAdd{amtrnn@gmail.com}

\abstract{
We investigate the thermal QCD phase transition and its scaling properties on the lattice.
The simulations are performed with $N_f=2+1+1$ Wilson twisted mass fermions at 
pion masses from physical up to heavy quark regime. We introduce a novel chiral order parameter,
which is free from linear mass contributions and turns out to be very useful for
the study of scaling behaviour. Our results are compatible with $O(4)$ universal scaling for the physical pion mass 
and the temperature range $[120:300]$ MeV. Violations to scaling at larger masses and other possible scenarios, 
including mean field behaviour and $Z(2)$ scaling scenario are also discussed. 
We provide an estimation for the critical temperature in the chiral limit $T_0$.
}

\FullConference{
 The 38th International Symposium on Lattice Field Theory, LATTICE2021
  26th-30th July, 2021
  Zoom/Gather@Massachusetts Institute of Technology
}

\begin{document}
\maketitle

\section{Introduction}

The nature of the thermal QCD phase transition as a function of quark masses has been the subject of numerous investigations (for example, see reviews~\cite{Ding:2020rtq,Guenther:2020jwe,Kotov:2021hri} and references therein). A~particularly interesting open question: {\it what are the universal properties of QCD phase transition in the limit of zero quark masses?} The most well-known arguments based on the $\epsilon$-expansion\cite{Pisarski:1983ms} predict that the thermal transition is of the first order for number of massless quark flavours $N_f \geq 3$. For $N_f=2$ the situation is less clear and depends on the fate of the axial $U_A(1)$ symmetry. If it remains broken at the critical temperature $T_c$, then the transition is of the second order with $O(4)$ universality class, while its effective restoration at the critical temperature would imply either another universality class~\cite{Pelissetto:2013hqa} or the first order phase transition. Nonzero quark masses explicitly violate chiral symmetry, and, in the case of the second order phase transition, the properties of observables for small quark mass should follow an universal scaling behaviour, in the so called scaling window around the phase transition. In the case of the first order phase transition, it should persist for small quark masses, ending in the $Z_2$ critical line. 

Despite substantial progress, the nature of the thermal QCD phase transition in the chiral limit and its scaling window remains an open issue. In this Proceeding we present the results of our study of the scaling behaviour of $N_f=2+1+1$ QCD in the limit of $N_f=2$ massless quarks. 

\section{Lattice details}

\begin{figure}[thb]
\begin{center}
\includegraphics[width=9cm]{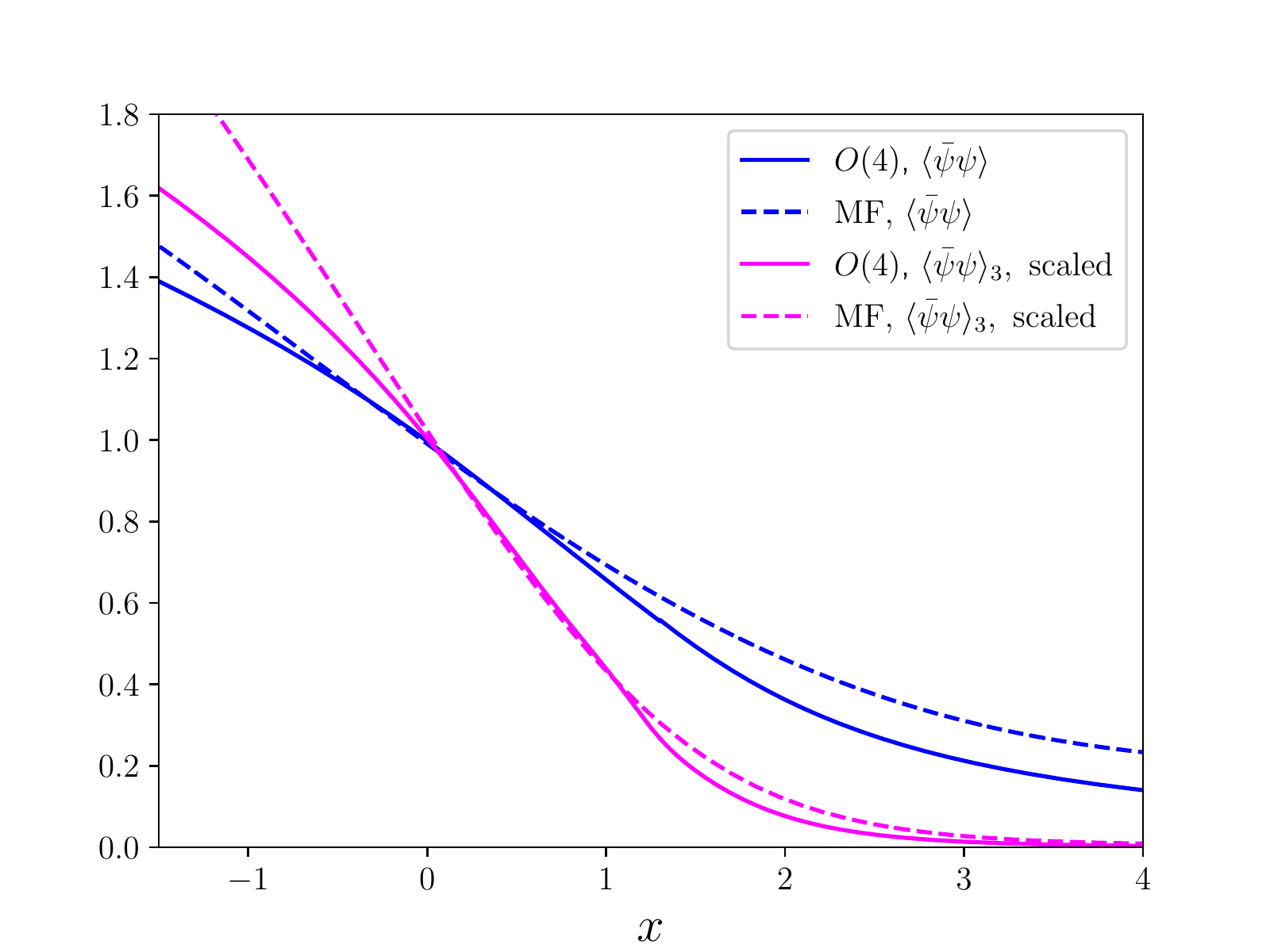}
\end{center}
\caption{Universal Equation of State for the chiral condensate and new order parameter $\langle\bar\psi\psi\rangle_3$ according to $O(4)$ universality class and mean field scaling. $\langle\bar\psi\psi\rangle_3$ is rescaled to go through $(0,1)$ point.}
\label{fig:universal_psibarpsi3}
\end{figure}

We perform simulations with $N_f=2+1+1$ twisted mass Wilson fermions at maximal twist~\cite{Frezzotti:2000nk}. Our study has been carried out in the fixed scale approach, where we keep lattice spacing $a$ fixed and change temperature by varying temporal lattice extent $N_t$. In the simulations the heavy quark sector was tuned to reproduce the masses of $K$ and $D$ mesons, and we performed the study of several ensembles for various pion masses.  Due to the tuning of parameters by ETM collaboration~\cite{Alexandrou:2018egz} we are able to do simulations with physical pion mass. A short summary of used ensembles is presented in Tab.~\ref{tab:ensembles}. Detailed discussion on lattice parameters and used statistics can be found in~\cite{Kotov:2021rah}. First preliminary results have been reported in~\cite{Kotov:2020hzm}.

\begin{table}[thb]
    \centering
    \begin{tabular}{|c|c|c|c|c|}
    \hline
         Ensemble & M140 & D210 & D370 & B370  \\
         \hline
         Pion mass $m_{\pi}$ [MeV] & 139.3(7) & 225(5) & 383(11) & 376(14) \\
         Lattice spacing $a$ [fm] & 0.0801(4) & 0.0619(18) & 0.0619(18) & 0.0815(30)\\
        \hline
    \end{tabular}
    \caption{Ensembles used in the study, corresponding pion mass and lattice spacing}
    \label{tab:ensembles}
\end{table}

In order to study thermal QCD phase transition and its scaling behaviour we investigated the following observables:
\begin{itemize}
    \item Chiral condensate $\langle\bar{\psi}\psi\rangle$
    \item Chiral susceptbility $\chi=\frac{\partial\langle\bar{\psi}\psi\rangle}{\partial m}$
    \item Combining chiral condensate and chiral susceptibility, we build a novel order parameter 
    \begin{equation}
    \langle\bar{\psi}\psi\rangle_3=\langle\bar{\psi}\psi\rangle-m\chi.
    \label{eq:psibarpsi3def}
    \end{equation}
    Note, that any linear in mass terms in the chiral condensate are cancelled in this observable, including regular nonuniversal terms and a leading order divergence $\sim m/a^2$.
\end{itemize}

\begin{figure}[thb]
\begin{center}
\includegraphics[width=4.9cm]{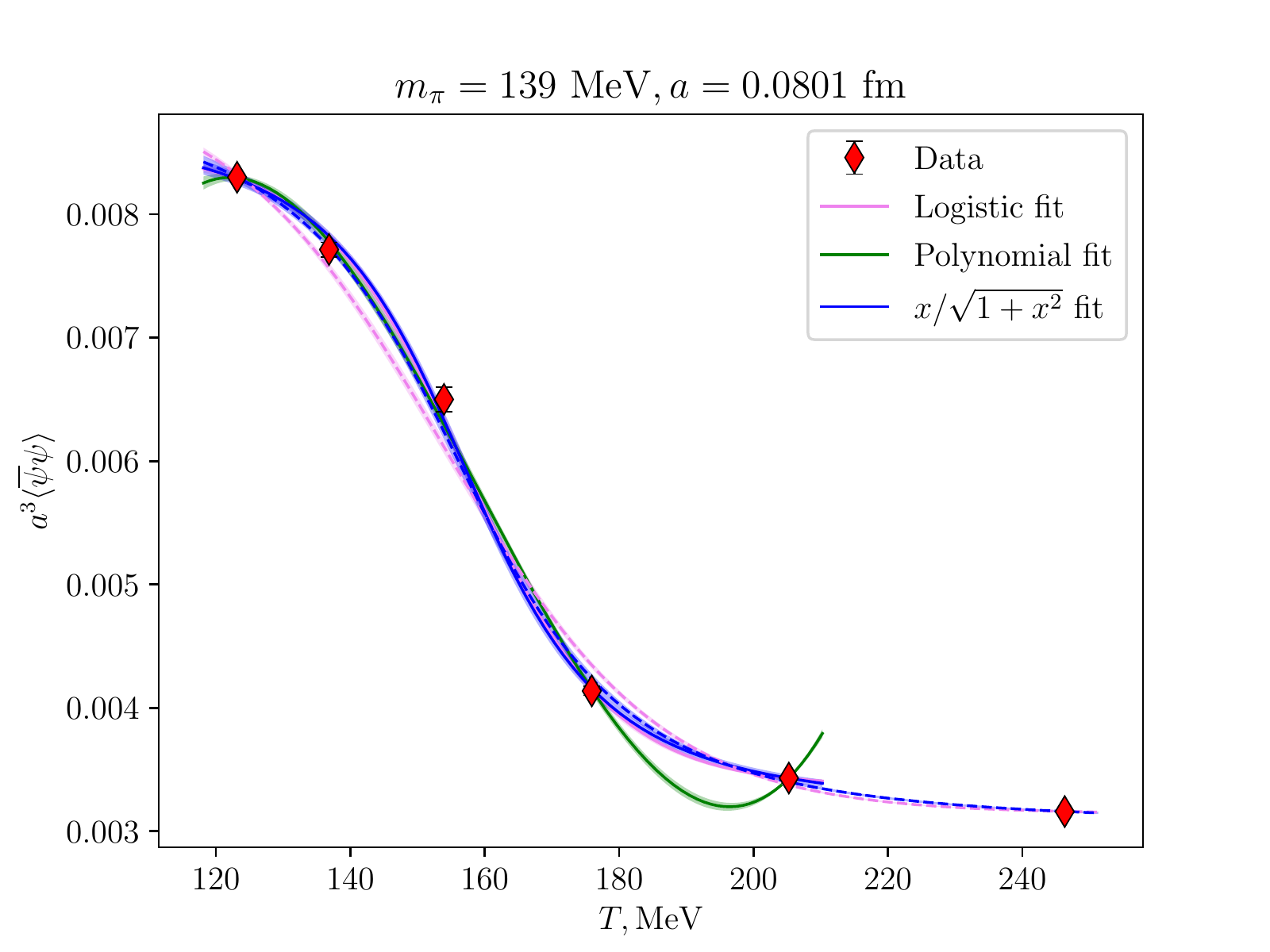}
\includegraphics[width=4.9cm]{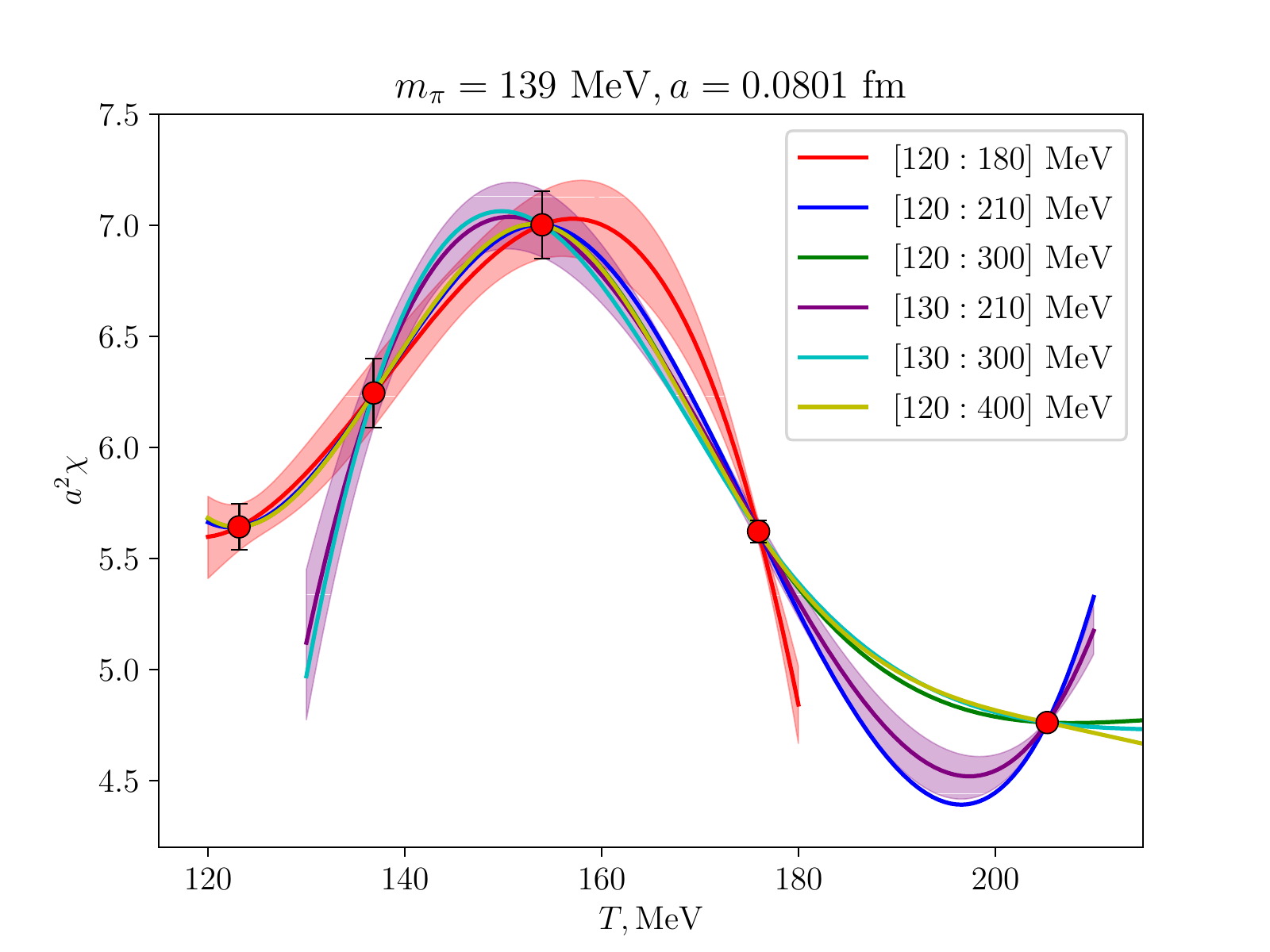}
\includegraphics[width=4.9cm]{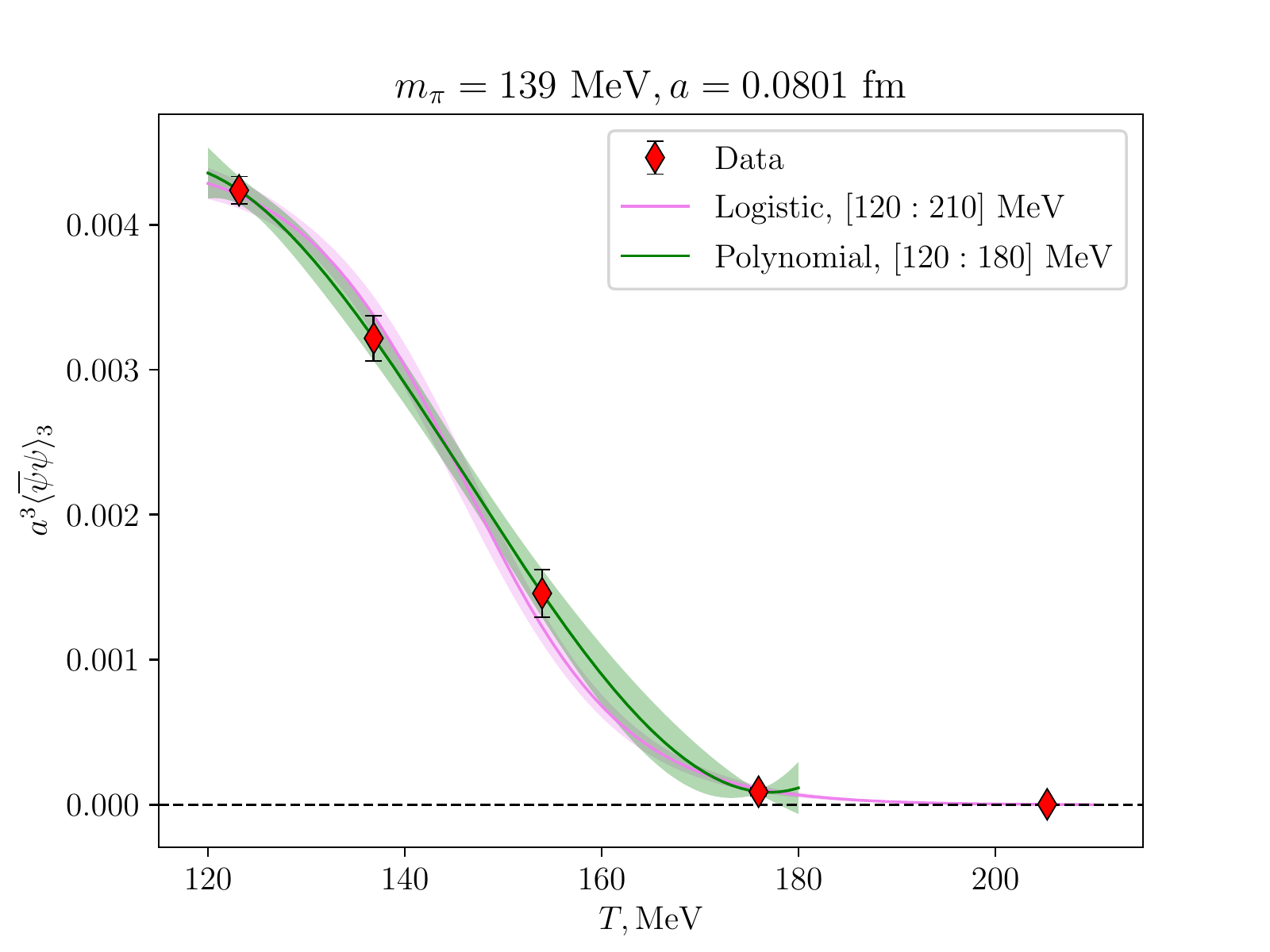}
\end{center}
\caption{Chiral condensate (left), chiral susceptibility (center) and new observable $\langle\bar{\psi}\psi\rangle_3$ (right) as functions of temperature for the physical pion mass $m_{\pi}=139$ MeV. For the chiral condensate and $\langle\bar{\psi}\psi\rangle_3$ we impose fits of various functional forms, for the chiral susceptibility a spline interpolation in various intervals is also presented. Figures from \cite{Kotov:2021rah}.}
\label{fig:observables_phys_pion}
\end{figure}

Taking the universal Equation of State for the chiral condensate:
$\langle\bar{\psi}\psi\rangle=m^{1/\delta}f(t/m^{1/\beta\delta})$, one can easily obtain for the new observable:
\begin{equation}
    \frac{\langle\bar{\psi}\psi\rangle_3}{m^{1/\delta}}=f(x)(1-1/\delta)+\frac{x}{\beta\delta}f'(x),
    \label{eq:eospsibarpsi3}
\end{equation}
here $t\equiv(T-T_0)$, $T_0$ is the critical temperature in the chiral limit, mass $m$ is an external symmetry breaking field, $x\equiv t/m^{1/\beta\delta}$, $\beta$ and $\delta$ are corresponding critical exponents.
The universal behaviour of $\langle\bar{\psi}\psi\rangle_3$ according to $O(4)$ universality class in comparison to the chiral condensate is presented in Fig.~\ref{fig:universal_psibarpsi3}. The pseudo-critical temperature extracted from the inflection point of $\langle\bar{\psi}\psi\rangle_3$ is given by $x=0.55(1)$ comparing to $x=0.74(4)$ from the inflection point of the chiral condensate, meaning that the pseudocritical temperature from this observable is closer to the critical temperature $T_0$ in the chiral limit. One should note much faster falloff of this observable in the large temperature region $\langle\bar\psi\psi\rangle_3 \propto t^{-\gamma - 2 \beta \delta}$
rather than $\langle \bar\psi\psi \rangle \propto  t^{-\gamma}$. These facts make the new observable very suitable for the study of the scaling behaviour.

\section{Results}

\begin{figure}[thb]
    \centering
    \includegraphics[width=12cm]{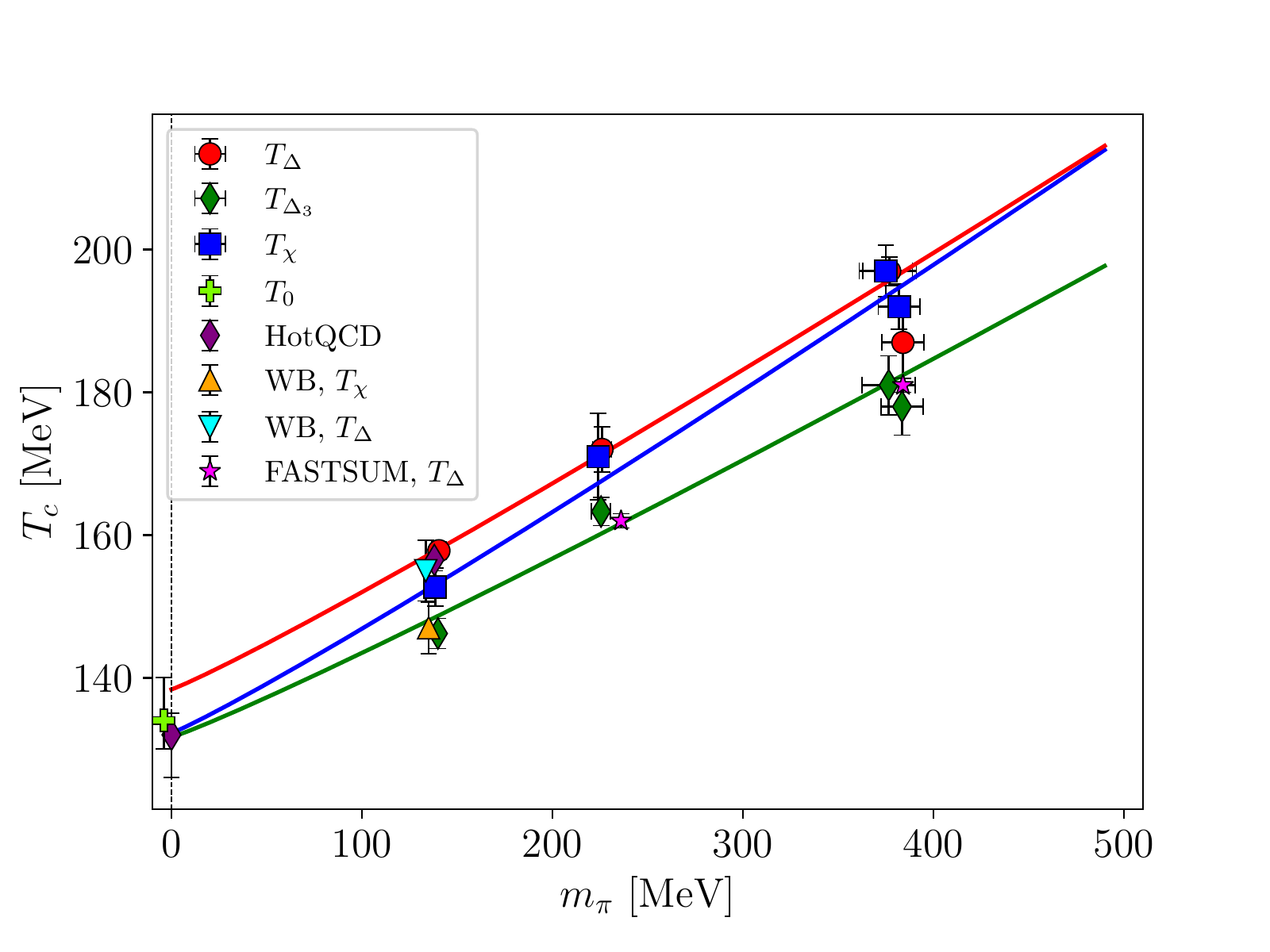}
    \caption{The pseudocritical temperature extracted from various observables as a function of pion mass $m_{\pi}$ together with $O(4)$ fits. Results of other groups: Budapest-Wuppertal~\cite{Borsanyi:2020fev} (orange and cyan triangles), HotQCD~\cite{Bazavov:2018mes} (purple diamonds) and FASTSUM~\cite{Aarts:2019hrg,Aarts:2020vyb} (violet stars) are also indicated. Light-green cross indicates our final estimation of the critical temperature $T_0$ in the chiral limit, Eq.~(\ref{eq:t_chiral_limit}). Points are slightly shifted for better readability. Figure from \cite{Kotov:2021rah}.}
    \label{fig:crittempvspion2}
\end{figure}

In Fig.~\ref{fig:observables_phys_pion} we present the dependence of the studied observables on the temperature for the physical pion mass. We fitted the chiral condensate and the new observable $\langle\bar\psi\psi\rangle_3$ with several functional forms in order to determine the inflection point and the pseudo-critical temperature. For the full chiral susceptibility, given small number of points and large regular contribution, we used cubic splines, which go through the points and their errorbars, to determine the peak position. We summarize the pseudo-critical temperatures for the physical pion mass in Tab.~\ref{tab:crittemp}. In Fig.~\ref{fig:crittempvspion2} we present the resulting dependence of the pseudo-critical temperature determined from all three observables as a function of pion mass, together with the results of other groups~\cite{Bazavov:2018mes,Aarts:2019hrg,Aarts:2020vyb, Borsanyi:2020fev}.  We fit the dependence of the pseudo-critical temperatures on the pion mass with the prediction of $O(4)$ universality class and determine the critical temperature in the chiral limit, which is also presented in Tab.~\ref{tab:crittemp}. Combining all observables, we obtain the following estimation of the pseudo-critical temperature in the chiral limit:
\begin{equation*}
    T_0=134^{+6}_{-4}\text{ MeV}.
    \label{eq:t_chiral_limit}
\end{equation*}

\begin{table}[]
    \centering
    \begin{tabular}{|c|c|c|c|}
    \hline
         Observable &  Chiral condensate & Chiral susceptibility &
         New order parameter $\langle\bar{\psi}\psi\rangle_3$\\
         \hline
         $T_c$ [MeV] for $m_{\pi}=m_{\pi}^{\text{phys}}$& 157.8(12) & 153(3) & 146(2)\\
         $T_c$ [MeV] for $m_{\pi}\to0$&138(2) & 132(4) & 132(3)\\
    \hline
    \end{tabular}
    \caption{Pseudocritical temperature for physical pion mass and critical temperature in the chiral limit, extracted from three studied observables.}
    \label{tab:crittemp}
\end{table}

\begin{figure}[thb]
\begin{center}
\includegraphics[width=12cm]{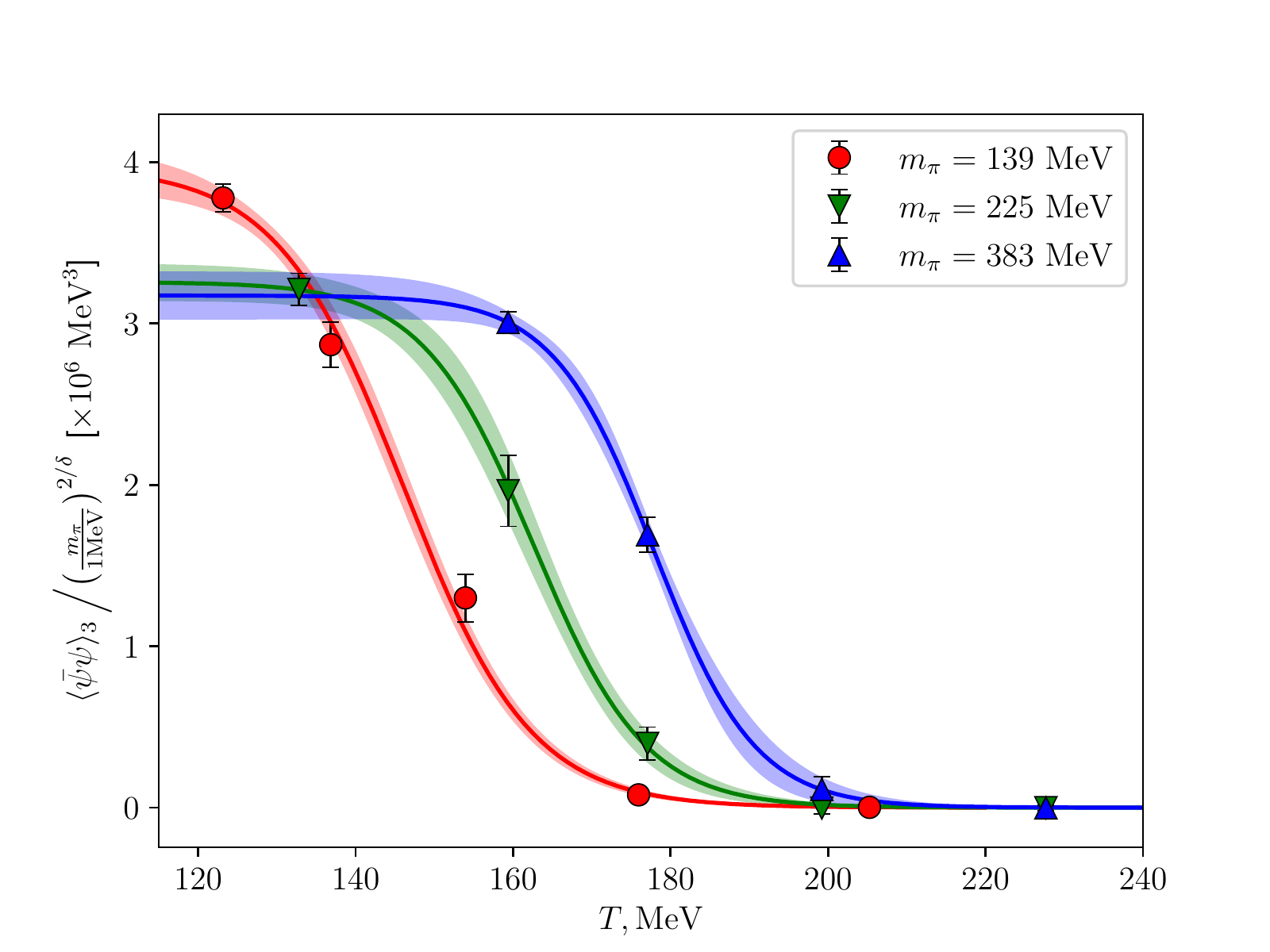}
\end{center}
\caption{New order parameter $\langle\bar\psi\psi\rangle_3$, divided by $m_{\pi}^{2/\delta}\sim m_q^{1/\delta}$. A crossing point of the curves for two lower pion masses provide an estimation of the critical temperature in the chiral limit $T_0=138(2)$ MeV according the EoS in Eq.~(\ref{eq:eospsibarpsi3}). Figure from \cite{Kotov:2021rah}.}
\label{fig:psibarpsi3rescaled}
\end{figure}

An alternative estimation of the critical temperature in the chiral limit can be obtained from the universal scaling of $\langle\bar\psi\psi\rangle_3$: if one takes the new observable $\langle\bar\psi\psi\rangle_3$ at the critical temperature in the chiral limit $t=0$, then, according to Eq.~(\ref{eq:eospsibarpsi3}), $\langle\bar\psi\psi\rangle_3/m^{1/\delta}\sim\langle\bar\psi\psi\rangle_3/m_{\pi}^{2/\delta}$ is indepenendent on the pion mass. In Fig.~\ref{fig:psibarpsi3rescaled} we plot the appropriately rescaled $\langle\bar\psi\psi\rangle_3$ for three different pion masses. From the intersection of the curves for two lower pion masses we obtain another estimation of the critical temperature in the chiral limit $T_0=138(2)$ MeV. Note, that this estimation is slightly higher, but consistent within errorbars with the estimation obtained from the extrapolation of the pseudocritical temperature, Eq.~(\ref{eq:t_chiral_limit}).

\begin{figure}
\centering
\begin{minipage}{.46\textwidth}
  \centering
  \includegraphics[width=7cm]{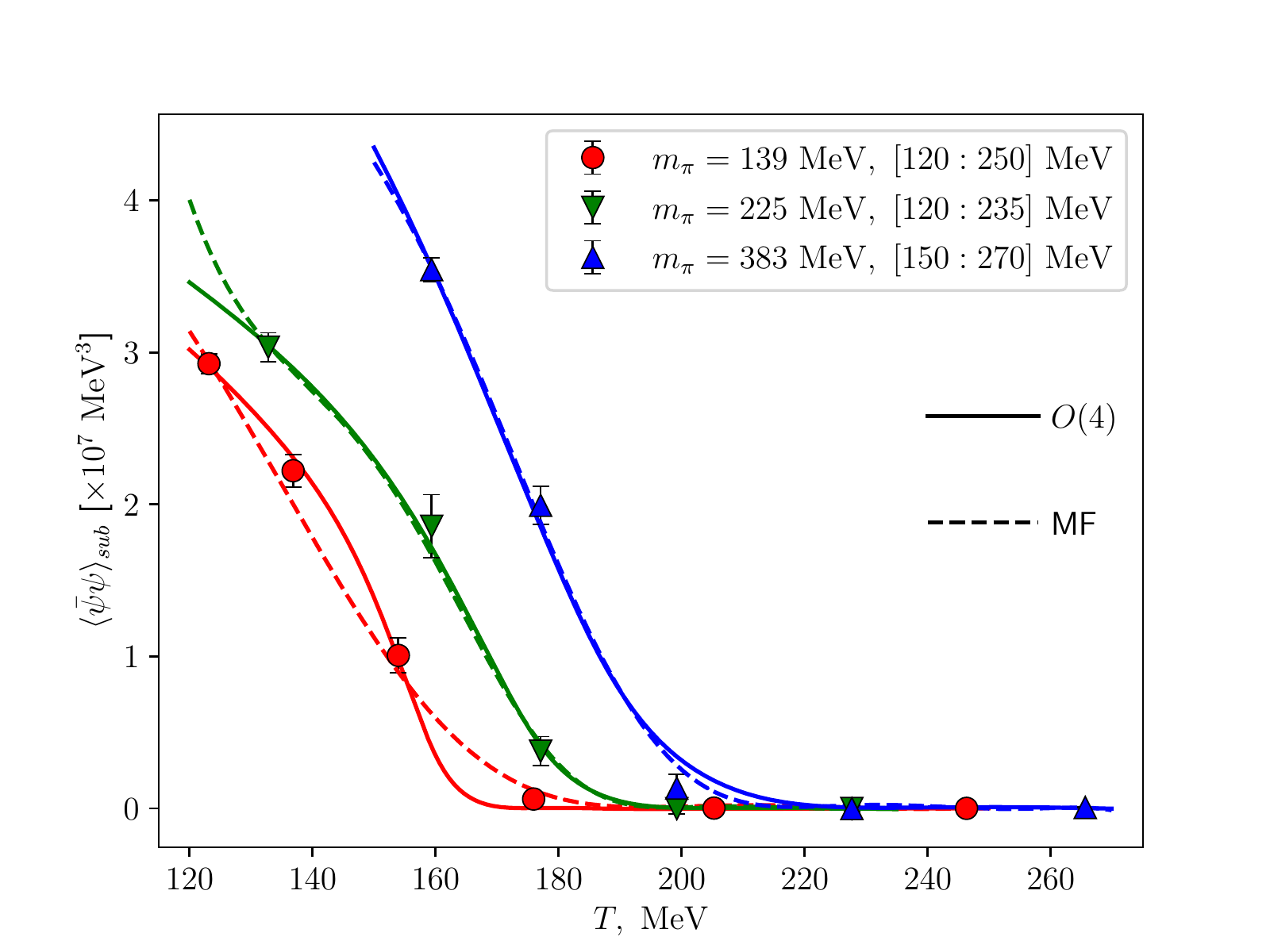}
  \captionof{figure}{Temperature dependence of $\langle\bar{\psi}\psi\rangle_3$ for three pion masses together with $O(4)$ and mean field fits. Figure from \cite{Kotov:2021rah}.}
  \label{fig:o4_mf_scaling}
\end{minipage}
\begin{minipage}{.05\textwidth}
~~
\end{minipage}
\begin{minipage}{.46\textwidth}
  \centering
  \includegraphics[width=7cm]{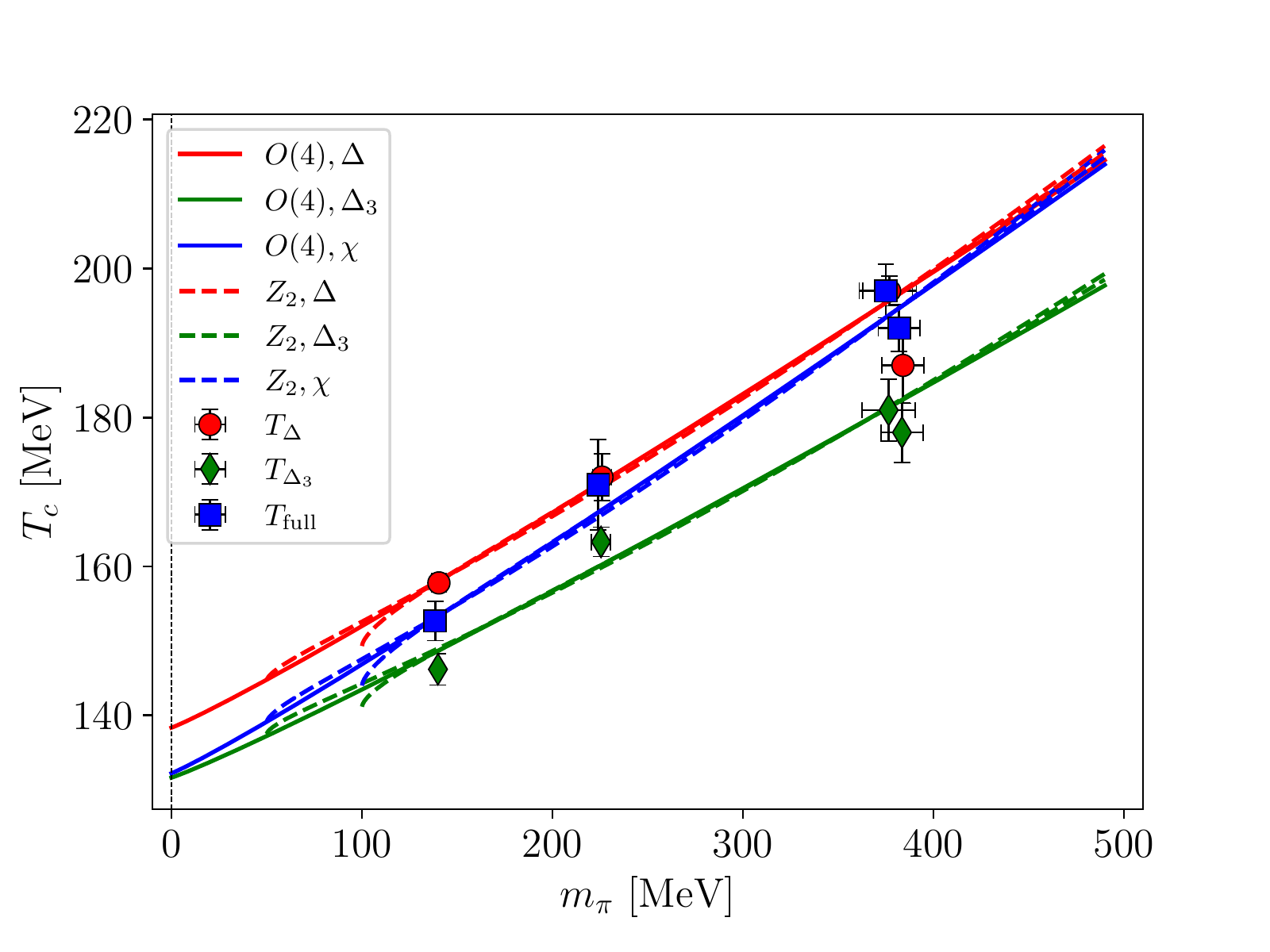}
  \captionof{figure}{The dependence of the pseudocritical temperatures on the pion mass together with fits, based on $O(4)$ predictions and $Z_2$ scaling with a critical pion mass $m_{\pi}^{cr}=50$ and $100$ MeV. Figure from \cite{Kotov:2021rah}.}
  \label{fig:temperature_o4_z2}
\end{minipage}
\end{figure}

It would be interesting to see, whether the dependence of the $\langle\bar{\psi}\psi\rangle_3$ on temperature can be described by a universal scaling behaviour. In Fig.~\ref{fig:o4_mf_scaling} we present this observable, together with the fits, given by $O(4)$ and mean field scaling. It can be seen that $O(4)$-based fits describe the data for all temperatures quite well. However, the critical temperature in the chiral limit determined from these fits is given by $T_0^{\text{EoS}}=142(2)$, 159(3), 174(2) MeV for pion masses  $m_{\pi}= 139$, 225 and 383 MeV correspondingly. The value of $T_0^{\text{EoS}}$ is close to the previous estimation (\ref{eq:t_chiral_limit}) of $T_0=134_{-4}^{+6}$~MeV only for the lightest pion. Note that mean field behaviour also describes our data, except for the lightest pion, where one sees some tension between our results and the fit. 

\begin{figure}[thb]
\begin{center}
\includegraphics[width=9cm]{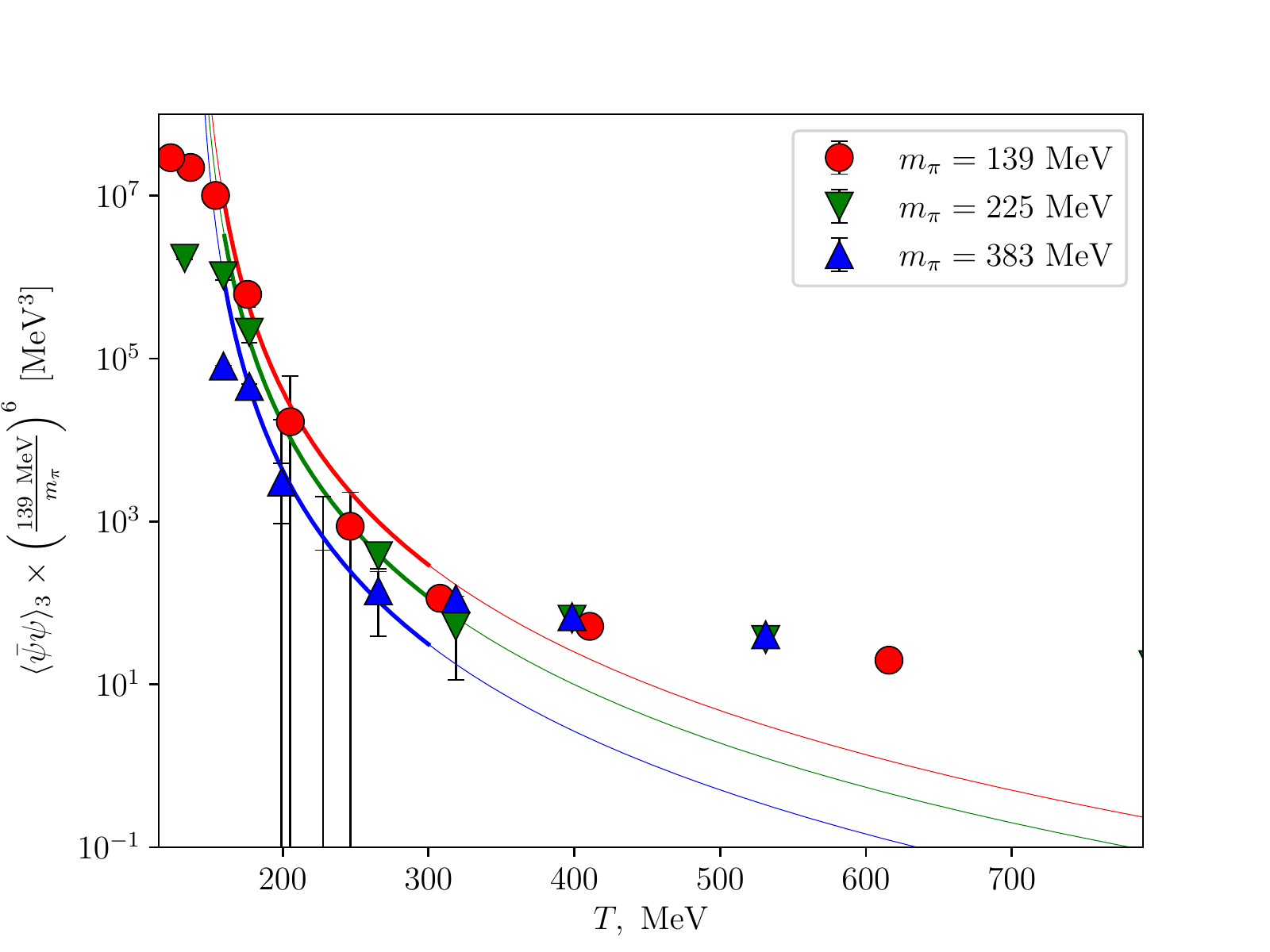}
\end{center}
\caption{Large temperature behaviour of the the rescaled $\langle\bar{\psi}\psi\rangle_3/m_q^3\sim\langle\bar{\psi}\psi\rangle_3/m_{\pi}^6$ together with the $O(4)$ fits. Fits work up to $\sim$ 300 MeV, at higher temperatures points for all pion mass fall on one curve, indicating a change to leading order Griffith analiticity $\langle\bar{\psi}\psi\rangle_3\sim m_q^3$. Figure from \cite{Kotov:2021rah}.}
\label{fig:larget}
\end{figure}

\begin{figure}[thb]
\begin{center}
\includegraphics[width=9cm]{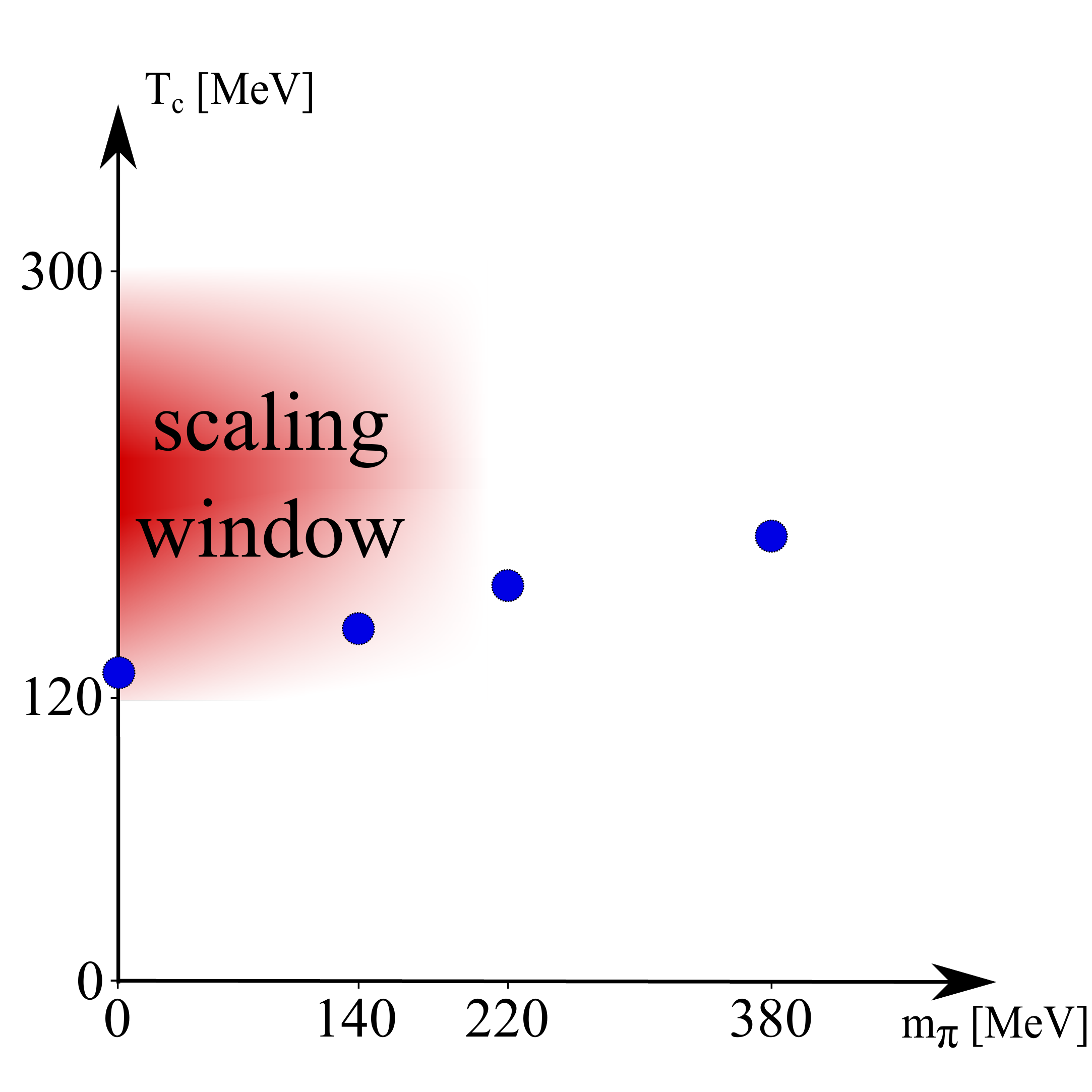}
\end{center}
\caption{Sketch of $O(4)$ scaling window in the temperature -- pion mass plane. Blue points indicate (pseudo)-critical temperatures for studied pion masses. Figure from \cite{Kotov:2021rah}.}
\label{fig:scalingwindow}
\end{figure}

We also checked possible $Z_2$ scaling. In Fig.~\ref{fig:temperature_o4_z2} we present the result of the fits of the pseudo-critical temperature versus pion mass with $Z_2$ behaviour with two fixed values of critical pion mass, $m_{\pi}^{\text{cr}}=50$ and $100$ MeV. Due to the closeness of critical exponents these fits are very close to the $O(4)$ predictions, and thus one cannot rule them out for any value of critical pion mass, $m_{\pi}^{\text{cr}}\lesssim 140$ MeV. 

Finally, in Fig.~\ref{fig:larget} we present the high temperature behaviour of the $\langle\bar{\psi}\psi\rangle_3$, divided by $m_{\pi}^6\sim m_q^3$. The $O(4)$ Equation of State predicts a fast falloff of this observable $\sim t^{-\gamma - 2 \beta \delta}$ after the phase transition. Indeed it is clearly seen, that the simple fit $\sim t^{-\gamma - 2 \beta \delta}$ nicely describes the data up to $\sim 300$ MeV. In the fit we have we used $T_0=138$ MeV -- the sensitivity to $T_0$ is rather moderate. At high temperatures $T\gtrsim 300$ MeV all points collapse into one curve, indicating a change to a leading order Griffith analyticity $\langle\bar{\psi}\psi\rangle_3\sim m_q^3$. Remarkably, in the study of topological susceptibility the same temperature $\sim300$~MeV was found to be a threshold for a dilute instanton gas behaviour~\cite{Burger:2018fvb,lat21_topology_proc}.

\section{Conclusions}

We have studied the properties of a thermal QCD phase transition with $N_f=2+1+1$ Wilson twisted mass fermions for several pion masses, starting from physical up to heavy quark regime. We introduced a novel order parameter $\langle\bar{\psi}\psi\rangle_3$, Eq. (\ref{eq:psibarpsi3def}), which turned out to very useful for the study of universal scaling behaviour of the phase transition. We determined the pseudo-critical temperature from the three different observables: chiral condensate, chiral susceptibility and a novel order parameter as functions of pion mass. Assuming $O(4)$ scaling, we extrapolated the pseudo-critical temperature to the chiral limit: $T_0=134_{-4}^{+6}$ MeV. Alternative method, based on the universal scaling of the novel observable  $\langle\bar{\psi}\psi\rangle_3$, gives slightly higher, although consistent within errorbars estimation $T_0=138(2)$ MeV. Closeness of the critical exponents does not allow to exclude possible $Z_2$ scaling of the critical temperatures with almost any critical pion mass $m_{\pi}^{\text{cr}}\lesssim 140$ MeV. The behaviour of $\langle\bar{\psi}\psi\rangle_3$ is consistent with $O(4)$ scaling window for pion masses $m_{\pi}\lesssim 140$ MeV and temperatures from 120 to 300 MeV, which we present in Fig.~\ref{fig:scalingwindow}. At temperatures $T>300$ MeV, the dependence changes to a leading order Griffith analyticity $\langle\bar{\psi}\psi\rangle_3\sim m_q^3$. 

\section*{Acknowledgements}

The work is partially supported by  STRONG-2020, a 
European Union’s Horizon 2020 research and innovation programme under grant agreement No. 824093.
A.Yu.K. and A.T. acknowledge support from RFBR grant 18-02-40126. A.T. was supported by the "BASIS" foundation.
Numerical simulations were carried out using computing resources of CINECA (based on the agreement between INFN and CINECA, on the ISCRA project IsB20), the supercomputer of Joint Institute for Nuclear Research ``Govorun'', and the computing resources of the federal collective usage center Complex for Simulation and Data Processing for Mega-science Facilities at NRC ``Kurchatov Institute'',~\url{http://ckp.nrcki.ru/}.

\bibliographystyle{JHEP}

\bibliography{biblio.bib}

\end{document}